\documentclass[aps,prl,twocolumn,superscriptaddress]{revtex4-1}
\usepackage{graphicx}% Include figure files
\usepackage{subfigure}
\usepackage{count1to}
\usepackage{dcolumn}% Align table columns on decimal point
\usepackage{bm}% bold math
\usepackage{amsmath}
\usepackage{color}
\usepackage{verbatim}
\usepackage{comment}
\usepackage{natbib} 
\usepackage{verbatim}
\usepackage{amsfonts}
\usepackage{lipsum}

\usepackage[colorlinks,
            linkcolor=blue,
            anchorcolor=blue,
            citecolor=blue,
urlcolor=blue
            ]{hyperref}
\begin{document}
\bibliographystyle{unsrt}

% Use the \preprint command to place your local institutional report
% number in the upper righthand corner of the title page in preprint mode.
% Multiple \preprint commands are allowed.
% Use the 'preprintnumbers' class option to override journal defaults
% to display numbers if necessary
%\preprint{}

%Title of paper
\title{Quantum-limited Localisation and Resolution in Three Dimensions}

% repeat the \author .. \affiliation  etc. as needed
% \email, \thanks, \homepage, \altaffiliation all apply to the current
% author. Explanatory text should go in the []'s, actual e-mail
% address or url should go in the {}'s for \email and \homepage.
% Please use the appropriate macro foreach each type of information

% \affiliation command applies to all authors since the last
% \affiliation command. The \affiliation command should follow the
% other information
% \affiliation can be followed by \email, \homepage, \thanks as well.
%\author{}
%\email[]{Your e-mail address}
%\homepage[]{Your web page}
%\thanks{}
%\altaffiliation{}
%\affiliation{}
\author{Ben Wang}
\affiliation{National Laboratory of Solid State Microstructures and Colloge of Engineering and Applied Sciences, Nanjing University, Nanjing 210093 China}
\author{Liang Xu}
\affiliation{National Laboratory of Solid State Microstructures and Colloge of Engineering and Applied Sciences, Nanjing University, Nanjing 210093 China}
\author{Jun-chi Li}
\affiliation{National Laboratory of Solid State Microstructures and Colloge of Engineering and Applied Sciences, Nanjing University, Nanjing 210093 China}
\author{Lijian Zhang}
\email{lijian.zhang@nju.edu.cn}
\affiliation{National Laboratory of Solid State Microstructures and Colloge of Engineering and Applied Sciences, Nanjing University, Nanjing 210093 China}
%Collaboration name if desired (requires use of superscriptaddress
%option in \documentclass). \noaffiliation is required (may also be
%used with the \author command).
%\collaboration can be followed by \email, \homepage, \thanks as well.
%\collaboration{}
%\noaffiliation

\date{\today}

% insert suggested PACS numbers in braces on next line
% insert suggested keywords - APS authors don't need to do this
%\keywords{}
\begin{abstract}
	As a method to extract information from optical systems, imaging can be viewed as a parameter estimation problem. The fundamental precision in locating one emitter or estimating the separation between two incoherent emitters is bounded below by the multiparameter quantum Cram\'{e}r-Rao bound (QCRB). Multiparameter QCRB gives an intrinsic bound in parameter estimation. We determine the ultimate potential of quantum-limited imaging for improving the resolution of a far-field, diffraction-limited optical field within the paraxial approximation. We show that the quantum Fisher information matrix (QFIm) about one emitter's position is independent on the true value of it. We calculate the QFIm of two unequal-brightness emitters' relative positions and intensities, and the results show that only when the relative intensity and centroids of two point sources including longitudinal and transverse direction are known exactly, the separation in different directions can be estimated simultaneously with finite precision. Our results give the upper bounds on certain far-field imaging technology and will find wide applications from microscopy to astrometry.
\end{abstract}
%\maketitle must follow title, authors, abstract, \pacs, and \keywords
\maketitle

% body of paper here - Use proper section commands
% References should be done using the \cite, \ref, and \label commands
\section{\expandafter{\romannumeral1}. Introduction}
% Put \label in argument of \section for cross-referencing
%\section{\label{}}
Locating an emitter and estimating different emitters' relative positions precisely are key tasks in imaging problems. 
The question of two-point resolution was first discussed by Rayleigh~\cite{r4,rayleigh1879xxxi}. Rayleigh's criterion states that two-point
sources are resolvable when the maximum of the illuminance
produced by one point coincides with the first
minimum of the illuminance produced by the other point. This criterion sets the limit of resolving power of optical systems~\cite{r4}. Many methods are developped to bypass this limit by converting resolving multi-emitter to locating single emitters. Deterministic super-resolution methods such as stimulated emission depletion (STED) microscopy \cite{Hell:94}, reversible saturable optical fluorescence transitions (RESOLFT) microscopy \cite{Heintzmann:02}, saturated structured illumination microscopy (SSIM) \cite{Gustafsson13081} utilize the fluorophores' nonlinear response to excitation, which leads to individual emitting of emitters. Stochastic super-resolution methods such as stochastic optical reconstruction microscopy (STORM) \cite{RN357} and photo-actived localization microscopy (PALM) \cite{Betzig1642} utilize the different temporal behavior of light sources, which emit light at separate times and thereby become resolvable in time. Therefore, localisation of a single emitter is also an essential and fundamental issue in imaging problems.

Imaging is, as its heart, a multiparameter problem \cite{PhysRevLett.123.143604}. Targets' localisation and resolution can be viewed as parameter estimation problems. Positions of emitters are treated as parameters encoded in quantum states. The minimal error to estimate these parameters is bounded by Cram\'{e}r-Rao lower bound (CRLB). To quantify the precision, researchers utilize Fisher information (FI)  associated with CRLB.

Inspired by classical and quantum parameter estimation theory~\cite{multiparameter,Matsumoto_2002,PhysRevLett.119.130504,PhysRevA.88.040102,vidrighin2014joint,PhysRevLett.111.070403,r15,r16,r13,giovannetti2011advances,PhysRevLett.96.010401},  Tsang and coworkers \cite{r5} reexamined Rayleigh's criterion. If only intensity is measured in traditional imaging, the CRLB tends to infinite as the separation between two point sources decreases, which is called Rayleigh curse. However, when the phase information is also taken into account, two incoherent point sources can be resolved no matter how close the separation is, which has been demonstrated in experiments~\cite{r6,PhysRevA.97.023830,Tsang_2017,PhysRevLett.118.070801,Bonsma_Fisher_2019}. If the centroid of the two emitters is also an unknown nuisance parameter, the precision to estimate the separation will decrease. Measuring the centroid precisely first can recover the lost precision due to misalignment between the measurement apparatus and the centroid \cite{grace2019approaching,r5}. Two-photon interference can be performed to estimate the centroid and separation at the same time \cite{PhysRevLett.121.250503}. Further developments in this emerging field have addressed the problem in estimating separation and centroid of two unequal brightness sources  \cite{mqmt,PhysRevA.98.012103,Prasad_2020}, locating more than two emitters  \cite{r11}, resolving the two emitters in three dimensional space \cite{r7,PhysRevLett.121.180504,PhysRevA.99.022116,r9,r18}, with partial coherence \cite{Larson:18,Tsang:19,Larson:19} and complete coherence \cite{Hradil:19}. In addition, with the development of super-resolution microscopy techniques mentioned above, the method to improve precision of locating a single emitter is also important. Efforts along this line include designing optimal point spread functions (PSF) \cite{r19,nehme2020learning} and the quantum-limited longitudinal localisation of a single emitter \cite{PhysRevLett.123.193601}.

\begin{figure}[h]
	\includegraphics[scale=0.3]{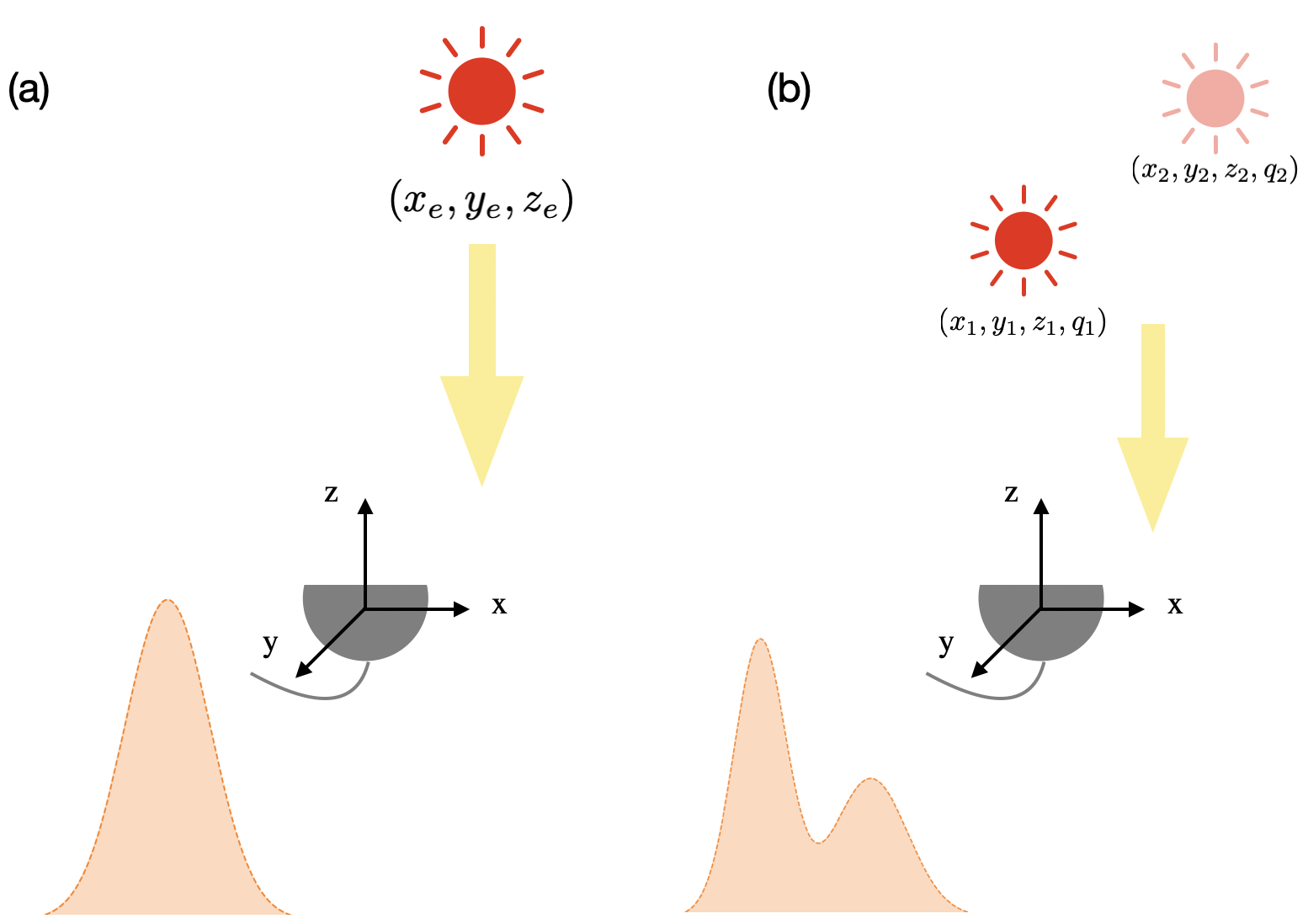}
	\caption{Schematic of one emitter with position ($x_0,y_0,z_0$) (a) and schematic of two emitters with positions ($x_1,y_1,z_1$) and ($x_2,y_2,z_2$), and different intensities ($q_1,q_2$) (b).}\label{fig:1}
\end{figure}
In this work, we generalize the quantum-limited super-resolution theory to the localisation of a single emitter with symmetric PSF and resolution of two unequal-brightness emitters in three dimensional space with arbitrary PSF. In the perspective of multiparameter estimation theory, we show that three Cartesian coordinates of single emitter's position (Fig. \ref{fig:1} (a)) can be estimated in a single measurement scheme. For two-emitter system, we consider the most general situation with five parameters including relative intensity, centroids, and separations in transverse and longitudinal direction, see Fig. \ref{fig:1} (b). We show that only two separations can be measured simultaneously to attain the quantum limit for the most general situation. In some special cases, centriods and separations can be estimated precisely at the same time. Localisation and resolution in three dimensions are important in both microscopy and astrometry. Our theoretical framework will be useful in these fields.

This paper is organised as follows: In Section {\uppercase\expandafter{\romannumeral2}}, we provide a quantum mechanical description of the optical system with one and two emitters; In Section \uppercase\expandafter{\romannumeral3}, we will review the quantum estimation theory, the main method to quantify the precision of localisation and resolution, and introduce the FI and quantum Fisher information (QFI). The specific expressions of QFI of localisation and resolution with some discussions will be provided in Section \uppercase\expandafter{\romannumeral4}, and some analysis will be done about the results. Finally, we summarize all the results in Section \uppercase\expandafter{\romannumeral5}.

\section{\expandafter{\romannumeral2}. QUANTUM DESCRIPTION OF LOCALISATION AND RESOLUTION}
 We assume that the emitters are point-like sources and the electromagnetic wave emitted by the emitters is quasimonochromatic and paraxial, with ($x,y$) denoting the image-plane coordinates, $z$ denoting the distance from the emitters to the image-plane. The quasimonochromatic paraxial wave  $\Psi(x-x_e,y-y_e,z_e)$ obeys the paraxial Helmholtz equation
\begin{equation}\label{1}
	\nabla^2_T\Psi+2k^2\Psi+i2k\frac{\partial}{\partial z}\Psi=0,
\end{equation}
where $(x_e,y_e,z_e)$ are unknown coordinates of the emitter with respect to the coordinate origin defined in the image-plane and 
 $\nabla^2_T\equiv\partial^2/\partial x^2+\partial^2/\partial y^2$. From Eq. (\ref{1}), the generator of the displacement in direction $z$ is $\hat{G}=\frac{1}{2k}\nabla^2_T+k$. The generators of the displacement in direction $x$ and $y$ are momentum operators $\hat{p}_x$ and $\hat{p}_y$, which are derivatives $-i\partial_x$ and $-i\partial_y$. We have $ \Psi(x-x_e,y-y_e,z_e)=\exp(-i\hat{G}z_e-i\hat{p}_xx_e-i\hat{p}_yy_e)\Psi(x,y,0)$. Then we rewrite the above results with quantum formulation and denote the PSF of the optical system $\Psi(x,y,0)=\langle x,y|\Psi\rangle$ with $|x,y\rangle=\hat{a}^\dagger(x,y)|0\rangle$. The quantum state of photons from a single emitter is
\begin{equation}\label{2}
	|\tilde{\Psi}\rangle=\exp(-i\hat{G}z_e-i\hat{p}_xx_e-i\hat{p}_yy_e)|\Psi\rangle,
\end{equation}
and here, $\tilde{\Psi}$ is the displaced wave function with respect to $\Psi(x,y,0)$.
\par
For two incoherent point sources, without the loss of generality, we only consider the displacement in $x$ and $z$ direction. The quantum state is
\begin{equation}\label{3}
	\rho=q|\Psi_1\rangle\langle\Psi_1|+(1-q)|\Psi_2\rangle\langle\Psi_2|,
\end{equation}
where $|\Psi_{1,2}\rangle=\exp(-i\hat{G}z_{1,2}-i\hat{p}_xx_{1,2})|\Psi\rangle$, and $(x_1,z_1)$ $(x_2,z_2)$ are coordinates of two incoherent light sources. Here, the relative intensity $q$ is also an unknown parameter. The density matrix $\rho$ gives the normalized mean intensity
\begin{equation}\label{4}
	\rho(x)=q|\Psi(x-x_1,z_1)|^2+(1-q)|\Psi(x-x_2,z_2)|^2.
\end{equation}
Eq. (\ref{3}) and Eq. (\ref{4}) can be reparameterized with the centroids $x_0\equiv(x_1+x_2)/2$, $z_0\equiv(z_1+z_2)/2$ and separations $s\equiv x_2-x_1$, $t\equiv z_2-z_1$. The parameter vector is $\boldsymbol{\theta}\equiv(x_0,dx,z_0,dz,q)^T$.

\section{\expandafter{\romannumeral3}. QUANTUM ESTIMATION THEORY}
Localisation and resolution can be treated as the estimation of the coordinates of  emitters. In this section, we review the quantum and classical estimation theory for further analysis. The quantum states in  both localisation and resolution problems are dependent on the parameters to be estimated. Let the parameters be $\boldsymbol{\theta} \equiv \{\theta_1,\theta_2,\theta_3,...\}^T$ and we use $\theta_i$ to substitute the parameters in Eq. (\ref{2}) and Eq. (\ref{3}) for convenience. A quantum measurement described by a positive operator-valued measure (POVM) $\Pi_j$ with the outcome $j$ is performed on the image plane to estimate $\boldsymbol{\theta}$, so that the probability distribution of the outcome is $p(j|\boldsymbol{\theta})=\operatorname{Tr}[\Pi_j\rho(\boldsymbol{\theta})]$. The estimators are $\check{\boldsymbol{\theta} }\equiv \{\check{\theta}_1,\check{\theta}_2,\check{\theta}_3,...\}^T$, which are the functions of measurement results. The precision of the estimates is quantified by the covariance matrix or mean square error
\begin{equation}\label{5}
	Cov[\boldsymbol{\theta}]\equiv\sum_{j}p(j|\boldsymbol{\theta})(\boldsymbol{\theta}-\check{\boldsymbol{\theta} }(j))^T(\boldsymbol{\theta}-\check{\boldsymbol{\theta} }(j)),
\end{equation}
$Cov(\boldsymbol{\theta})$ is a positive symmetric matrix with diagonal element denoting the variances of each estimator. The non-diagonal elements denote the covariance between different estimators.
\par
For unbiased estimators, the covariance matrix is lower bounded by the Cram\'{e}r-Rao bound
\begin{equation}\label{6}
	Cov[\boldsymbol{\theta}]\ge\frac{1}{M}[F(\rho_{\boldsymbol{\theta}},\Pi_j)]^{-1},
\end{equation}
where M is the number of copies of the system to obtain the estimators $\check{\boldsymbol{\theta}}$. 
$F(\rho_{\boldsymbol{\theta}},\Pi_j)$ is the Fisher information matrix (FIm) defined by
\begin{equation}\label{7}
[F(\rho_{\boldsymbol{\theta}},\Pi_j)]_{\mu\nu}=\sum_{j}\frac{1}{p(j|\boldsymbol{\theta})}\frac{\partial p(j|\boldsymbol{\theta})}{\partial\theta_\mu}\frac{\partial p(j|\boldsymbol{\theta})}{\partial\theta_\nu},
\end{equation}
where $\mu$ and $\nu$ denote the row and column index of the FIm. Inequality in Eq. (\ref{6}) means the matrix $	Cov[\boldsymbol{\theta}]-\frac{1}{M}[F(\rho_{\boldsymbol{\theta}},\Pi_j)]^{-1}$ is semi-positive definite matrix.
\par
Here, we give an example of FIm that the measurement method is the intensity detection, projecting the quantum state into the eigenstates of the spatial coordinates. The elements of this POVM are $\left\{\Pi_{x,y}=|x,y\rangle\langle x,y|\right\}$, and the FIm
\begin{equation}\label{8}
F^{direct}_{\mu\nu}=\int\int\frac{1}{p(x,y|\boldsymbol{\theta})}\frac{\partial p(x,y|\boldsymbol{\theta})}{\partial\theta_\mu}\frac{\partial p(x,y|\boldsymbol{\theta})}{\partial\theta_\nu}dxdy,
\end{equation}
with $p$($x,y$)=Tr($\rho \Pi_{x,y}$).
\par
To get the ultimate precision, it is necessary to get the bound which only depends on the quantum states rather than the measurement systems
\begin{equation}\label{9}
		Cov[\boldsymbol{\theta}]\ge\frac{1}{M}[F(\rho_{\boldsymbol{\theta}},\Pi_j)]^{-1}\ge\frac{1}{M}[Q(\rho_{\boldsymbol{\theta}})]^{-1},
\end{equation}
where the $Q(\rho_{\boldsymbol{\theta}})$ is the quantum Fisher information matrix (QFIm) which gives the maximum FIm. Its matrix elements are given by
\begin{equation}\label{10}
[Q(\rho_{\boldsymbol{\theta}})]_{\mu\nu}=\frac{1}{2}\operatorname{Tr}[\rho_{\boldsymbol{\theta}}\{L_\mu ,L_\nu\}],
\end{equation}
in which $\{\cdot,\cdot\}$ denotes anticommutator, and $L_\kappa$ stands for the symmetric logarithmic derivative (SLD) with respect to the parameter $\theta\kappa$, which satisfies the condition
\begin{equation}\label{11}
	\partial_\kappa\rho_{\boldsymbol{\theta}}=\frac{L_\kappa\rho_{\boldsymbol{\theta}}+\rho_{\boldsymbol{\theta}}L_\kappa}{2}.
\end{equation}
\par
For multiparameter estimation problem, an essential issue is the attainability of QCRB. If the system only has a single parameter to be estimated, the optimal measurement is to project the quantum state onto the eigenstates of the SLD \cite{r13}, while this strategy is not suitable for mutiple parameters. If the SLD operators $L_\kappa$ corresponding to the different parameters commute with each other $([L_\mu,L_\nu]=0)$, there exists a measurement which can maximize the parameters' estimation precision simultaneously. If not, it does not imply this bound can not be saturated. As discussed in \cite{Matsumoto_2002,r15,r16}, a sufficient and necessary condition for the saturability of the QCRB in Eq. (\ref{9}) is the satisfaction of weak commutativity condition
\begin{equation}\label{12}
	\operatorname{Tr}[\rho_{\boldsymbol{\theta}}[L_\mu ,L_\nu]]=0.
\end{equation}
We define the weak commutativity condition matrix $\Gamma(\rho_{\boldsymbol{\theta}})$, and $[\Gamma(\rho_{\boldsymbol{\theta}})]_{\mu\nu}=\frac{1}{2i}\operatorname{Tr}[\rho_{\boldsymbol{\theta}}[L_\mu ,L_\nu]]$.
\section{\expandafter{\romannumeral4}. RESULTS}
\par
Our main results contain two parts. First, we show the QFIm of locating an emitter with symmetric wave functions satisfying paraxial Helmholtz equation in three dimensional space. Second, we give the QFIm of two incoherent point sources in which the parameters to be estimated include relative intensitiy, centroids and separations in both transverse and longitudinal direction.
\par
\section{A. QUANTUM localisation IN THREE DIMENSIONAL SPACE}
In general, we assume that the wave function is symmetric in transverse direction with respect to its center
\begin{equation}\label{13}
\Psi(x,y,z)=\Psi(-x,y,z)=\Psi(x,-y,z).
\end{equation}
Considering the situation of a single emitter, the quantum state is a pure state in Eq. (\ref{2}). The SLD can be written in the simple expression
\begin{equation}\label{14}
	L_\kappa=2(|\tilde{\Psi}\rangle\langle\partial_\kappa\tilde{\Psi}|+|\partial_\kappa\tilde{\Psi}\rangle\langle\tilde{\Psi}|),
\end{equation}
where $|\partial_\kappa\tilde{\Psi}\rangle=\partial|\tilde{\Psi}\rangle/\partial\theta\kappa$. Moreover, since $\partial_\kappa\langle\tilde{\Psi}|\tilde{\Psi}\rangle=\langle\partial_\kappa\tilde{\Psi}|\tilde{\Psi}\rangle+\langle\tilde{\Psi}|\partial_\kappa\tilde{\Psi}\rangle=0,$ QFIm can be written in the form
\begin{equation}\label{15}
	\left[Q_{loc}(\boldsymbol{\theta})\right]_{j k}=4 \operatorname{Re}(\langle\partial_{j} \tilde{\Psi}| \partial_{k} \tilde{\Psi}\rangle-\langle\partial_{j} \tilde{\Psi} | \tilde{\Psi}\rangle\langle\tilde{\Psi} | \partial_{k} \tilde{\Psi}\rangle),
\end{equation}
where Re denotes the real part. The specific forms of $|\partial_\kappa\tilde{\Psi}\rangle$ in this problem are
\begin{equation}\label{16}
\begin{aligned}
&|\partial_{x_e}\tilde{\Psi}\rangle=-i\hat{p}_x|\tilde{\Psi}\rangle,\\
&|\partial_{y_e}\tilde{\Psi}\rangle=-i\hat{p}_y|\tilde{\Psi}\rangle,\\
&|\partial_{z_e}\tilde{\Psi}\rangle\ =-i\hat{G}|\tilde{\Psi}\rangle,
\end{aligned}
\end{equation}
because of the symmetry of the wave function in Eq. (\ref{13}), $\langle\tilde{\Psi} | \partial_{k} \tilde{\Psi}\rangle=-\langle\Psi|\partial_\kappa|\Psi\rangle=0$ for any $\kappa=x,y$. The weak commutativity condition is
\begin{equation}\label{17}
	\left[\Gamma_{loc}(\boldsymbol{\theta})\right]_{j k}=4 \operatorname{Im}(\langle\partial_{j} \tilde{\Psi}| \partial_{k} \tilde{\Psi}\rangle-\langle\partial_{j} \tilde{\Psi} | \tilde{\Psi}\rangle\langle\tilde{\Psi} | \partial_{k} \tilde{\Psi}\rangle),
\end{equation}
where Im denotes the imaginary part. According to the Eq. (\ref{15}) and (\ref{16}), we get the QFIm
\begin{equation}\label{18}
	Q_{loc}=4\left[
	\begin{matrix}
	p_x^2&0&0\\0&p_y^2&0\\0&0&g_z^2-G_z^2
	\end{matrix}
	\right],
\end{equation}
with $p_x=\sqrt{\langle\Psi|\hat{p}_x^2|\Psi\rangle}$, $p_y=\sqrt{\langle\Psi|\hat{p}_y^2|\Psi\rangle}$, $g_z=\sqrt{\langle\Psi|\hat{G}^2|\Psi\rangle}$ and $G_z=\langle\Psi|\hat{G}|\Psi\rangle$. The weak commutativity condition is satisfied since
\begin{equation}\label{19}
\Gamma_{loc}=\left[
\begin{matrix}
0&0&0\\0&0&0\\0&0&0
\end{matrix}
\right].
\end{equation}
This result indicates that the 3D localisation problem is compatible \cite{r16}, i.e., we can perform a single measurement to estimate all the parameters simultaneously and attain the precision achieved by optimal measurement for each parameter.
If the generators for each parameters commute with each other $[\hat{G}_i,\hat{G}_j]=0$, the weak commutativity condition is always satisfied.  This is indeed the situation for the generators $\hat{p}_x,\hat{p}_y$ and $\hat{G}$.
\par
We take the Gaussian beam as an example, which is the most common beam in practical experiments. 
The pure state without displacement in Eq. (\ref{2}) is
\begin{equation}
|\Psi\rangle=\int_{x,y}dxdy\sqrt{\frac{2}{\pi w_0^2}}\exp\left(-\frac{x^2+y^2}{w_0^2}\right)|x,y\rangle,
\end{equation}
with $w_0$ the waist radius.
\begin{figure}[ht]
	\includegraphics[scale=0.3]{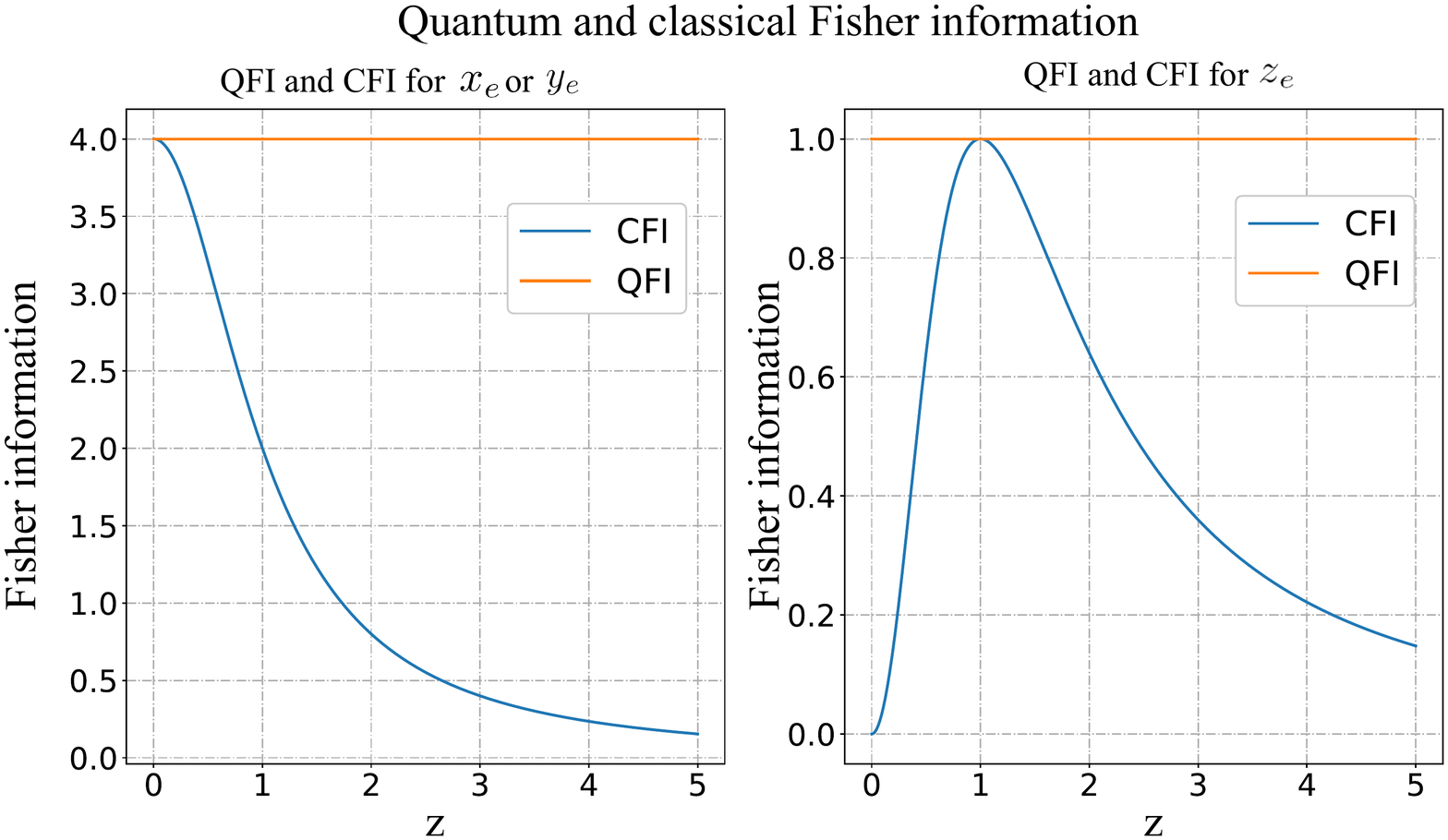}
	\setlength{\abovecaptionskip}{-1pt}
	\setlength{\belowcaptionskip}{-1pt}
	\caption{Quantum and classical Fisher information of localisation in three dimensional space. For the estimation of the transverse coordinates of the emitter, the CFI coincides with the QFI in the position $z=0$, which indicates intensity measurement achieves QFI if the detector is put in the position of waist. While for the estimation of the longitudinal coordinate, the detector needs to be put at the Rayleigh rangle to get the best precsion.\label{fig:2}}
\end{figure}
The shifted wave function is
\begin{equation}\label{gaussian}
\begin{aligned}
|\tilde{\Psi}\rangle=&\int_{x,y}dxdy\sqrt{\frac{2}{\pi w(z_e)^2}}\exp\left(-\frac{(x-x_e)^2+(y-y_e)^2}{w(z_e)^2}\right)\\&\exp\left(-ikz_e-\! ik\frac{(x-x_e)^2+(y-y_e)^2}{2R(z_e)}\!+i\zeta(z_e)\right)|x,y\rangle,
\end{aligned}
\end{equation}
with $\!w(z_e)=w_0\sqrt{1\!+\!(z_e/z_r)^2}\!\!$ ,  $\!R(z_e)=z_e\! \left[1\!+\! (z_r/z_e)^2\right]\!\!\!$ and $\zeta(z_e)=\tan^{-1}(z_e/z_r)$, where $z_r$ is the Rayleigh range of a Gaussian beam which equals to $
\pi w_{0}^2/\lambda$ related to the wavelength $\lambda$.
 
The result of QFIm is
\begin{equation}
		4\left[
	\begin{matrix}
	\frac{1}{w_0^2}&0&0\\0&\frac{1}{w_0^2}&0\\0&0&\frac{1}{4z_r^2}
	\end{matrix}
	\right].
\end{equation}
Considering the conventional intensity measurement, the classical Fisher information (CFI), according to Eq. (\ref{8}), is
\begin{equation}
F_{\mu\nu}=\int_{x,y}dxdy\frac{1}{I(x,y)}\frac{\partial I(x,y)}{\partial\theta_{\mu}}\frac{\partial I(x,y)}{\partial\theta_{\nu}},
\end{equation}
with $I(x,y)=|\langle x,y|\tilde{\Psi}\rangle|^2$, the CFIs of three parameters are
\begin{equation}
\begin{aligned}
&F_{x_ex_e}=\frac{4z_r^2}{w_0^2(z^2+z_r^2)},\\
&F_{y_ey_e}=\frac{4z_r^2}{w_0^2(z^2+z_r^2)},\\
&F_{z_ez_e}=\frac{4z^2}{(z^2+z_r^2)^2}.
\end{aligned}
\end{equation}
From these results, we can see that if only intensity measurement is applied when the detector is at the position of waist, the CFIs for $x_e$ and $y_e$ equal to the QFIs, while in $z$ direction, the detector should be put at the Rayleigh range. Estimation of different parameters requires us to put the detector at different positions, which indicates that the intensity measurement is not the optimal measurement. The optimal measurement methods remain to be explored.  To improve the precision of estimation, we can optimize the input state. Shaping the wave function to change the PSFs of optical systems is also helpful here \cite{r17,r18,r19}. Another beam often used in experiments is Laguerre–Gauss (LG) beam.  Recent work shows the precision to estimate longitudinal position using LG beam is better than Gaussian beam \cite{Koutn2021}. We also calculate the QFI of transverse position of LG beam, and show the ratio between the QFI of Gaussian beam and that of LG beam in the table (\ref{table}) with respect to the azimuthal mode index $\boldsymbol{p}$ and radial index $\boldsymbol{l}$. The results show using LG beam to locate an emitter's transverse position also has a better performance than Gaussian.

\begin{table}[ht]
	\begin{tabular}{|c|c|c|c|c|}
		\hline $QFI_{LG}/QFI_{G}$& $\boldsymbol{p}=0$ & $\boldsymbol{p}=1$ & $\boldsymbol{p}=2$ & $\boldsymbol{p}=3$ \\
		\hline $|\boldsymbol{l}|=0$ & 1 & 3 & 5 & 7 \\
		\hline $|\boldsymbol{l}|=1$ & 2 & 4 & 6 & 8 \\
		\hline $|\boldsymbol{l}|=2$ & 3 & 5 & 7 & 9 \\
		\hline $|\boldsymbol{l}|=3$ & 4 & 6 & 8 & 10 \\
		\hline
	\end{tabular}
\caption{Ratio between the QFI of Gaussian beam and that of LG beam with respect to the azimuthal mode index $\boldsymbol{p}$ and radial index $\boldsymbol{l}$. Here, we select $\boldsymbol{p}$=0,1,2,3, and $\boldsymbol{l}$=0,1,2,3. ($\boldsymbol{p}$,$\boldsymbol{l}$)=(0,0) is the Gaussian beam.}\label{table}
\end{table}

\par
\section{B. QUANTUM LIMITED RESOLUTION IN THREE DIMENSIONS}
\par
Now we consider two incoherent point sources with the quantum state in Eq. (\ref{3}). Different from single emitters, the quantun state is a mixed state, which implies Eq. (\ref{15}) can not be used here. we need a new method to calculate the QFIm. According to the definition of SLD in Eq. (\ref{11}), we find the quantum state $\rho$ and its derivatives which is associated with SLDs are supported in the subspace spanned by $|\psi_1\rangle$, $|\psi_2\rangle$, $\partial_{x_1}|\psi_1\rangle$, $\partial_{z_1}|\psi_1\rangle$, $\partial_{x_2}|\psi_1\rangle$ and $\partial_{z_2}|\psi_1\rangle$. Thus,
similar to \cite{r9}, our analysis relies on the expansion of the quantum state $\rho$ in the non-orthogonal but normalized basis,
\begin{equation}\label{non}
	\{|\Psi_1\rangle,|\Psi_2\rangle,|\Psi_3\rangle,|\Psi_4\rangle,|\Psi_5\rangle,|\Psi_6\rangle\},
\end{equation}
where 
\setlength\abovedisplayskip{0pt}
\setlength\belowdisplayskip{0pt}
\begin{widetext}
\begin{equation}\label{ba}
\begin{aligned}
&|\Psi_1\rangle=\exp(-i\hat{G}z_1-i\hat{p}x_1)|\Psi\rangle,\ \ \ \ \ \ \ |\Psi_2\rangle=\exp(-i\hat{G}z_2-i\hat{p}x_2)|\Psi\rangle,\\
&|\Psi_3\rangle=\frac{-i\hat{p}\ \exp(-i\hat{G}z_1-i\hat{p}x_1)|\Psi\rangle}{\mathfrak{p}}, |\Psi_4\rangle=\frac{-i\hat{G}\ \exp(-i\hat{G}z_1-i\hat{p}x_1)|\Psi\rangle}{\mathfrak{g}},\\
&|\Psi_5\rangle=\frac{-i\hat{p}\ \exp(-i\hat{G}z_2-i\hat{p}x_2)|\Psi\rangle}{\mathfrak{p}}, |\Psi_6\rangle=\frac{-i\hat{G}\ \exp(-i\hat{G}z_2-i\hat{p}x_2)|\Psi\rangle}{\mathfrak{g}},\\
\ \
\end{aligned}
\end{equation}
\end{widetext}
with $\mathfrak{p}\!=\!\sqrt{\langle\Psi|\hat{p}^2|\Psi\rangle}$, $\mathfrak{g}\!=\! \sqrt{\langle\Psi|\hat{G}^2|\Psi\rangle}$. The relation between the representation of quantum states based on orthogonal basis and non-orthogonal basis is linear transformation shown in appendix.
The derivation of QFIm and weak commutativity condition matrix is also shown in appendix.
After a lengthy calculation, we get the two matrices
\begin{widetext}
\begin{equation}\label{38}
Q=\left[\begin{matrix}
Q_{x_0x_0}&2 \mathfrak{p}^{2}(1-2 q)&Q_{x_0z_0}&0&4w \partial_{s} w \\
2 \mathfrak{p}^{2}(1-2 q)&\mathfrak{p}^2&0&0&0\\Q_{x_0z_0}&0&Q_{z_0z_0}&2\left(\mathfrak{g}^{2}-\mathfrak{G}^{2}\right)(-1+2 q)&4 w\partial_{t} w \\0&0&2\left(\mathfrak{g}^{2}-\mathfrak{G}^{2}\right)(-1+2 q)&\mathfrak{g}^{2}-\mathfrak{G}^{2}&0\\
4 w\partial_{s} w&0&4w \partial_{t} w &0&\frac{-1+w^{2}}{(-1+q) q}
\end{matrix}\right],
\end{equation}
\begin{equation}\label{39}
\Gamma=\left[\begin{matrix}
0&\Gamma_{x_0s}&\Gamma_{x_0z_0}&\Gamma_{x_0t}&4 \partial_{s} \phi(-1+2 q) w^{2} \\
-\Gamma_{x_0s}&0&\Gamma_{sz_0}&0&-2 \partial_{s} \phi w^{2}\\-\Gamma_{x_0z_0}&-\Gamma_{sz_0}&0&\Gamma_{z_0t}&4\left(\mathfrak{G}+\partial_{t} \phi\right)(-1+2 q) w^{2}\\-\Gamma_{x_0t}&0&-\Gamma_{z_0t}&0&-2\left(\mathfrak{G}+\partial_{t} \phi\right) w^{2}\\
-4 \partial_{s} \phi(-1+2 q) w^{2}&2 \partial_{s} \phi w^{2}&-4\left(\mathfrak{G}+\partial_{t} \phi\right)(-1+2 q) w^{2} &2\left(\mathfrak{G}+\partial_{t} \phi\right) w^{2}&0
\end{matrix}\right],
\end{equation}
where 
\begin{equation}\label{40}
\begin{aligned}
&we^{i\phi}=\langle\Psi_1|\Psi_2\rangle,\\
&\mathfrak{G}\!=\!\langle \Psi|\hat{G}|\Psi\rangle,\\
&Q_{x_0x_0}=4\left(\mathfrak{p}^{2}-4\left(\partial_{s} w\right)^{2}(1-q) q-\frac{4\left(\partial_{s} \phi\right)^{2}(1-q) q w^{2}}{1-w^{2}}\right),\\
&Q_{x_0z_0}=16 \partial_{s} w \partial_{t} w(-1+q) q-\frac{16 \partial_{s} \phi\left(\mathfrak{G}+\partial_{t} \phi\right)(-1+q) q w^{2}}{-1+w^{2}},\\
&Q_{z_0z_0}=\frac{4(\mathfrak{G}^{2}-4\left(\partial_{t} w\right)^{2}(-1+q) q-\left(\mathfrak{G}^{2}-4\left(\mathfrak{G}-\partial_{t} w+\partial_{t} \phi\right)\left(\mathfrak{G}+\partial_{t} w+\partial_{t} \phi\right) q(1-q)\right)w^2)}{-1+w^2}+4\mathfrak{g}^2,\\
&\Gamma_{x_0s}=-\frac{8\partial_{s} w \partial_{s} \phi(-1+q) q w^{3}}{-1+w^{2}},\\
&\Gamma_{x_0z_0}=-16\left(-\partial_{s} \phi \partial_{t} w+\partial_{s} w\left(\mathfrak{G}+\partial_{t} \phi\right)\right)(-1+q) q(-1+2 q) w,\\
&\Gamma_{x_0t}=-\frac{8(-1+q) q w\left(\partial_s\phi\partial_t w+\partial_s w(\mathfrak{G}+\partial_t \phi)(-1+w^2)\right)}{-1+w^{2}},\\
&\Gamma_{sz_0}=-\frac{8(-1+q) q w\left(\partial_{s} w\left(\mathfrak{G}+\partial_{t} \phi\right)+\partial_{s} \phi \partial_{t} w\left(-1+w^{2}\right)\right)}{-1+w^{2}},\\
&\Gamma_{z_0t}=-\frac{8 \partial_{t} w\left(\mathfrak{G}+\partial_{t} \phi\right)(-1+q) q w^{3}}{-1+w^{2}}.\\
\end{aligned}
\end{equation}
\end{widetext}

If the separation in longitudinal direction is zero and the centroid in this direction is known, matrix in Eq.  (\ref{38}) reduces to a 3$\times$3 matrix, same to the result of Ref. \cite{mqmt}. If the wave function satisfies the equation
\begin{equation}\label{cond}
	 \mathfrak{G}+\partial_t\phi=0,
\end{equation}
   the parameters $z_0$, $t$ and $q$ can be estimated with the precision given by QCRB simultaneously. In the most general case, for an arbitrary wave function, only the separations in $x$ and $z$ directions satisfy the weak commutativity condition. Therefore, the QFIm becomes
\begin{equation}\label{41}
\left[
\begin{matrix}
\mathfrak{p}^2&0\\0&\mathfrak{g}^2-\mathfrak{G}^2
\end{matrix}
\right],
\end{equation}
in which each element is a constant.
In brief, parameters on separations in $x$ and $z$ direction are compatible. In multiparameter estimation problem, the achievable precision bound is Helovo Cram\'{e}r-Rao bound (HCRB) \cite{helstrom1976quantum,holevo2011probabilistic}, denoted by $c_h$. The discrepancy $\mathfrak{D}$ between QCRB and HCRB which equals to $c_h-\operatorname{Tr}(Q^{-1})$  is bounded by \cite{Carollo_2019}
\begin{equation}\label{ineq}
	0\leq \mathfrak{D} \leq \operatorname{Tr}(Q^{-1})\mathfrak{R},
\end{equation}
with $\mathfrak{R}:=\left\|i\Gamma Q^{-1}\right\|_\infty$, where $\left\|\cdot \right\|_\infty$ is the largest eigenvalue of a matrix. The first inequality is saturated iff Eq. (\ref{12}) is satisfied. $\mathfrak{R}$ is a quantitative indicator of compatibility in multiparameter estimation problems whose value is between 0 and 1 \cite{Carollo_2019}. Eq. (\ref{ineq}) shows that if $\mathfrak{R}$ equals to zero, HCRB equals to QCRB. Meanwhile, HCRB is at most twice QCRB \cite{Carollo_2019,PhysRevX.10.031023}.

\begin{figure*}[ht]
	\centerline{\includegraphics[scale=0.8]{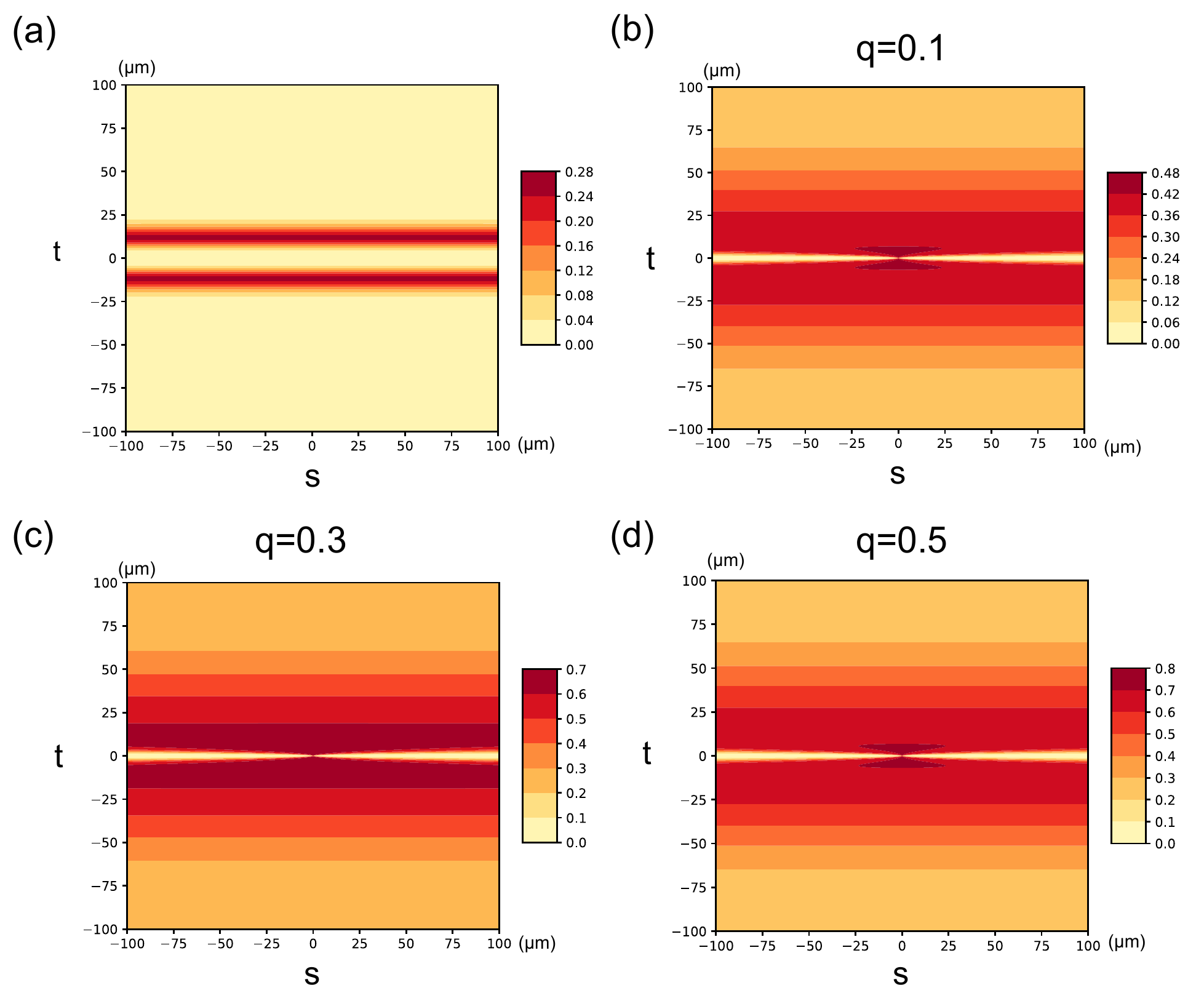}}
	\setlength{\abovecaptionskip}{-1pt}
	\setlength{\belowcaptionskip}{-1pt}
	\caption{Contour plot of $\mathfrak{R}$ of two Gaussian incoherent beams model in three dimensions in the ($t,s$) plane. (a) Relative intensity is a constant and equals to 0.5. (b) Relative intensity is also a parameter to be estimated, here we set $q=0.1$. (c) Similar to (b) while $q=0.3$. (d) Similar to (b) while $q=0.5$. \label{fig:3}}
\end{figure*}

We take the Gaussian beam in Eq. (\ref{gaussian}) as an example. We obtain $\mathfrak{p}=\frac{1}{w_0}$, $\mathfrak{g}=\sqrt{k^2+\frac{2}{k^2w_0^4}-\frac{2}{w_0^2}}$, $\mathfrak{G}=k-\frac{1}{kw_0^2}$, $w=\sqrt{\frac{1}{1+(\frac{t}{2z_r})^2}}\exp(-\frac{kz_rs^2}{t^2+4z_r^2})$, $\phi=\tan^{-1}(\frac{t}{2z_r})-kt(1+\frac{s^2}{2t^2+8z_r^2})$
and $\mathfrak{g}^2-\mathfrak{G}^2=1/kw_0^4$. Condition (\ref{cond}) is satisfied iff $t=0$. Here, the value of $\mathfrak{R}$ is shown in Fig. \ref{fig:3} with $w_0=100 \ \mu m$, and wavelength $\lambda=0.5\  \mu m$. In Fig. \ref{fig:3} (a), the relative intensity is a constant $q=0.5$, while in the other three pictures, relative intensity is also a parameter to be estimated. From these results, we find $\mathfrak{R}$ is close to zero in some regions, especially when the separations in two directions are nearly zero.

When the separations in $x$ and $z$ direction is infinitesimal (far less than the wavelength), the QFIm $Q_G$ and weak commutativity condition matrix $\Gamma_G$ of the Gaussian beam become 

\begin{equation}\label{43}
\lim_{s,t\rightarrow 0}Q_{G}=\left[
\begin{matrix}
\frac{2k}{z_r}&\frac{k(1-2q)}{z_r}&0&0&0\\\frac{k(1-2q)}{z_r}&\frac{k}{2z_r}&0&0&0\\0&0&\frac{1}{z_r^2}&\frac{-1+2q}{2z_r^2}&0\\0&0&\frac{-1+2q}{2z_r^2}&\frac{1}{4z_r^2}&0\\0&0&0&0&0
\end{matrix}
\right],
\end{equation}
and

\begin{equation}\label{44}
\lim_{s,t\rightarrow 0}\Gamma_{G}=\left[
\begin{matrix}
0&0&0&0&0\\0&0&0&0&0\\0&0&0&0&0\\0&0&0&0&0\\0&0&0&0&0
\end{matrix}
\right],
\end{equation}
indicating that except the intensity, the other four parameters can be estimated simultaneously and the optimal precision of each parameter is a constant. Different intensities of the two emitters introduce the statistical correlations between the separation and centroid in the same direction. The parameters in different directions have negligible correlation even though the intensities of two point sources are different. Off-diagonal terms of QFIm lead to the inequality, $[Q(\rho_{\boldsymbol{\theta}})^{-1}]_{jj}\ge 1/Q(\rho_{\boldsymbol{\theta}})_{jj} $, which means the existence of off-diagonal terms reduce the precision to estimate each parameter. Meanwhile, different intensities and the separation in longitudinal direction arise the asymmetry of two point sources, which reduces the precision to estimate the centroids in both transverse and longitudinal direction. Compared to Ref.\cite{r9}, our results analyse how different intensities affect the four parameters in transverse and longitudinal direction, and here, relative intensity is also considered as an unknown parameter to be estimated.
These results may find applications in sub-wavelength imaging.

\section{\expandafter{\romannumeral5}. CONCLUSION AND DISCUSSION}

In summary, we give the general model and fundamental limitation for the localisation of a single emitter and resolution of two emitters in three dimensional space. For one emitter, although the parameters in three directions are compatible with each other, the intensity detection can not extract the maximal information of three-dimensional positions simultaneously. Optimal measurement methods remain to be explored.

For two emitters, there are five parameters including the relative intensity, separations and centroids in transverse and longitudinal direction of two emitters. We have obtained the quantum-limited resolution via the QFIm. In the most general case that one do not have any prior information of these parameters, only separations in longitudinal and transverse direction can be estimated simultaneously to achieve the quantum-limited precision. More parameters can achieve the quantum-limited precision under some special conditions like Eq. (\ref{cond}). The example of Gaussian beam shows that if and only if separation in longitudinal direction is zero, one can estimate separation, centroid in longitudinal direction and the relative intensity with the quantum-limited precision. The example also shows that when the separations in two directions are much smaller than the wavelength, all of the elements in the QFIm are constants, which indicates that separations and centroids in longitudinal and transverse directions can be estimated precisely with a single measurement scheme.
Spatial-mode demultiplexing \cite{r6,PhysRevA.97.023830,Tsang_2017} or mode sorter \cite{r7} can be useful here.

We should note that our results is suitable not only for Gaussian beams, but also for arbitrary symmetric wave functions satisfying paraxial Helmholtz equation. Our results give a fundamental bound of quantum limit in localisation and resolution in the three dimensional space and will stimulate the development of new imaging methods.

\section{APPENDIX: SPECIFIC FORMULATIONS OF THE DERIVATIVE OF QUANTUM STATE}\label{app} 
\setcounter{equation}{0}
 In this appendix, we give the derivation of QFIm and weak commutativity condition matrix. From Eq. (\ref{3}) and Eq. (\ref{ba}), we have 
\begin{equation}\label{rho}\tag{A1}
\begin{aligned}
&\rho|\Psi_j\rangle=q\Pi_{1j}|\Psi_1\rangle+(1-q)\Pi_{2j}|\Psi_2\rangle,
\end{aligned}
\end{equation}
where $\Pi_{ij}=\langle\Psi_i|\Psi_j\rangle$.
Therefore, $\rho$ can be expressed as a matrix form
\begin{widetext}
	\begin{equation}\label{qfi}\tag{A2}
	R=\left[\begin{matrix}
	q\Pi_{11}&q\Pi_{12}&q\Pi_{13}&q\Pi_{14}&q\Pi_{15}&\Pi_{15}\\(1-q)\Pi_{21}&(1-q)\Pi_{22}&(1-q)\Pi_{23}&(1-q)\Pi_{24}&(1-q)\Pi_{25}&(1-q)\Pi_{26}\\0&0&0&0&0&0\\0&0&0&0&0&0\\0&0&0&0&0&0\\0&0&0&0&0&0
	\end{matrix}\right],
	\end{equation}
	
\end{widetext}
It is non-hermitian because we use the non-orthogonal basis. By Gram-Schimidt process, we can get the orthonormal basis $\{|e_1\rangle,|e_2\rangle,|e_3\rangle,|e_4\rangle,|e_5\rangle,|e_6\rangle\}$  and the matrix ($\rho$) in this basis is similar to matrix (\ref{qfi}) which means $\rho=TRT^{-1}$, where $T$ is the transformation matrix between the orthonormal basis $\{|e_i\rangle,i=1,...,6\} $ and non-orthogonal basis mentioned in Eq. (\ref{non}) The same method can be used to get the expressions of $\partial_{\theta_i}\rho$:

\setlength\abovedisplayskip{0pt}
\setlength\belowdisplayskip{0pt}
\begin{equation}\label{derivative}\tag{A3}
\begin{aligned}
&\partial_{x_1}\rho=q\mathfrak{p}(|\Psi_3\rangle\langle\Psi_1|+|\Psi_1\rangle\langle\Psi_3|),\\
&\partial_{x_2}\rho=(1-q)\mathfrak{p}(|\Psi_5\rangle\langle\Psi_2|+|\Psi_2\rangle\langle\Psi_5|),\\
&\partial_{z_1}\rho=q\mathfrak{g}(|\Psi_4\rangle\langle\Psi_1|+|\Psi_1\rangle\langle\Psi_4|),\\
&\partial_{z_2}\rho=(1-q)\mathfrak{g}(|\Psi_6\rangle\langle\Psi_2|+|\Psi_2\rangle\langle\Psi_6|),\\
&\partial_q\rho=|\Psi_1\rangle\langle\Psi_1|-|\Psi_2\rangle\langle\Psi_2|.\\
\end{aligned}
\end{equation}
The specific formulations of these matrices are shown in Appendix. Then, to get the QFIm of two emitters, it is necessary to solve the Eq. (\ref{11}) to get the SLDs of different parameters,
\begin{equation}\label{32}\tag{A4}
\Xi_{\theta_i}=\frac{R\mathbb{L}_{\theta_i}+\mathbb{L}_{\theta_i}R}{2},
\end{equation}
where $\Xi_{\theta_i}$ is the matrix representation of $\partial_{\theta_i}$ under the non-orthogonal basis, where
\begin{equation}\label{A1}\tag{A5}
\Xi_{x_1}=qp\left[\begin{matrix}
\Pi_{31}&\Pi_{32}&\Pi_{33}&\Pi_{34}&\Pi_{35}&\Pi_{36}\\0&0&0&0&0&0\\\Pi_{11}&\Pi_{12}&\Pi_{13}&\Pi_{14}&\Pi_{15}&\Pi_{16}\\0&0&0&0&0&0\\0&0&0&0&0&0\\0&0&0&0&0&0
\end{matrix}\right],
\end{equation}

\begin{equation}\tag{A6}
\Xi_{x_2}=(1-q)p\left[\begin{matrix}
0&0&0&0&0&0\\\Pi_{51}&\Pi_{52}&\Pi_{53}&\Pi_{54}&\Pi_{55}&\Pi_{56}\\0&0&0&0&0&0\\0&0&0&0&0&0\\\Pi_{21}&\Pi_{22}&\Pi_{23}&\Pi_{24}&\Pi_{25}&\Pi_{26}\\0&0&0&0&0&0
\end{matrix}\right],
\end{equation}
\begin{equation}\tag{A7}
\Xi_{z_1}=qg\left[\begin{matrix}
\Pi_{41}&\Pi_{42}&\Pi_{43}&\Pi_{44}&\Pi_{45}&\Pi_{46}\\0&0&0&0&0&0\\0&0&0&0&0&0\\\Pi_{11}&\Pi_{12}&\Pi_{13}&\Pi_{14}&\Pi_{15}&\Pi_{16}\\0&0&0&0&0&0\\0&0&0&0&0&0
\end{matrix}\right],
\end{equation}

\begin{equation}\Xi_{z_2}=(1-q)g\left[\begin{matrix}\tag{A8}
0&0&0&0&0&0\\\Pi_{61}&\Pi_{62}&\Pi_{63}&\Pi_{64}&\Pi_{65}&\Pi_{66}\\0&0&0&0&0&0\\0&0&0&0&0&0\\\Pi_{21}&\Pi_{22}&\Pi_{23}&\Pi_{24}&\Pi_{25}&\Pi_{26}\\0&0&0&0&0&0
\end{matrix}\right],
\end{equation}
and 
\begin{equation}\tag{A9}
\begin{matrix}
\Xi_{q}=\left[\begin{matrix}
\Pi_{11}&\Pi_{12}&\Pi_{13}&\Pi_{14}&\Pi_{15}&\Pi_{16}\\-\Pi_{21}&-\Pi_{22}&-\Pi_{23}&\-\Pi_{24}&-\Pi_{25}&-\Pi_{26}\\0&0&0&0&0&0\\0&0&0&0&0&0\\0&0&0&0&0&0\\0&0&0&0&0&0
\end{matrix}\right].
\end{matrix}
\end{equation}

Since estimating the separation and centroid of two point sources is equivalent to estimating the position of each emitter, we can use new parameters $(x_0,s,z_0,t)$ to replace the previous four $(x_1,x_2,z_1,z_2)$, and relative intensity remains unchanged.

\begin{equation}\label{33}\tag{A10}
\begin{aligned}
&\theta_1=x_0=\frac{x_2+x_1}{2},\theta_2=s=x_2-x_1,\\
&\theta_3=z_0=\frac{z_2+z_1}{2},\theta_4=t=z_2-z_1,\\
&\theta_5=q.
\end{aligned}
\end{equation}

The relation between the SLDs of the new parameters with respect to the old ones can be written as

\begin{equation}\label{relation}\tag{A11}
\left(\begin{array}{c}{\hat{L}_{x_{0}}} \\ {\hat{L}_{s}} \\ {\hat{L}_{z_{0}}} \\ {\hat{L}_{t}} \\ {\hat{L}_{q}}\end{array}\right)=\left(\begin{array}{ccccc}{1} & {1} & {0} & {0} & {0} \\ {-\frac{1}{2}} & {\frac{1}{2}} & {0} & {0} & {0} \\ {0} & {0} & {1} & {1} & {0} \\ {0} & {0} & {-\frac{1}{2}} & {\frac{1}{2}} & {0} \\ {0} & {0} & {0} & {0} & {1}\end{array}\right)\left(\begin{array}{c}{\hat{L}_{x_{1}}} \\ {\hat{L}_{x_{2}}} \\ {\hat{L}_{z_{1}}} \\ {\hat{L}_{z_{2}}} \\ {\hat{L}_{q}}\end{array}\right).
\end{equation}

Now, we take the SLD of $x_0$ as an example to show the relation in Eq. (\ref{relation}). The parameter $x_0$ has the same generator $\hat{p}$ with $x_1$ and $x_2$. According to Eq. (\ref{3}) and Eq. (\ref{ba}), $\partial_{x_1}\rho=iq\left[|\Psi_1\rangle\langle\Psi_1|,\hat{p}\right]$,  $\partial_{x_2}\rho=i(1-q)\left[|\Psi_2\rangle\langle\Psi_2|,\hat{p}\right]$ and 
\begin{equation}\tag{A12}
\begin{aligned}
&\left|\Psi_{1}\right\rangle=\exp \left(-i \hat{G} z_{1}-i \hat{p} x_{1}\right)|\Psi\rangle\\&=\exp \left(-i \hat{G} z_{1}-i \hat{p} (x_{0}-\frac{s}{2})\right)|\Psi\rangle,\\
&\left|\Psi_{2}\right\rangle=\exp \left(-i \hat{G} z_{2}-i \hat{p} x_{2}\right)|\Psi\rangle\\&=\exp \left(-i \hat{G} z_{2}-i \hat{p} (x_{0}+\frac{s}{2})\right)|\Psi\rangle.\\
\end{aligned}
\end{equation}
So $\partial_{x_0}|\Psi_1\rangle=\partial_{x_1}|\Psi_1\rangle$ and $\partial_{x_0}|\Psi_2\rangle=\partial_{x_2}|\Psi_2\rangle$, then we can get

\begin{equation}\label{exa}\tag{A13}
\partial_{x_0}\rho=i\left[\rho,\hat{p}\right]=\partial_{x_1}\rho+\partial_{x_2}\rho.
\end{equation}
From the definition of SLD Eq. (\ref{11}) and Eq. (\ref{exa}), we can show that
\begin{equation}
\hat{L}_{x_0}=\hat{L}_{x_1}+\hat{L}_{x_2}.
\end{equation}
The other relations of SLDs can be derived in a similar way.

Next, QFIm and weak commutativity condition matrix can be derivated from Eq. (\ref{10}) and Eq. (\ref{12})
\begin{equation}\label{36}\tag{A14}
\left[Q\left(\rho\right)\right]_{\mu \nu}+i\left[\Gamma\left(\rho\right)\right]_{\mu \nu}=\operatorname{Tr}\left[\rho L_{\mu} L_ \nu\right],
\end{equation}
\setlength\abovedisplayskip{0pt}
where

\setlength\abovedisplayskip{0pt}
\setlength\abovedisplayskip{0pt}
\begin{equation}\label{37}\tag{A15}
\begin{aligned}
\operatorname{Tr}\left[\rho L_{\mu} L_ \nu\right]&=\operatorname{Tr}[TRT^{-1}T\mathbb{L}_\mu T^{-1}T\mathbb{L}_\nu T^{-1}]\\
&=\operatorname{Tr}[R\mathbb{L}_\mu \mathbb{L}_\nu].
\end{aligned}
\end{equation}
\par

$Note\ added $. we are aware of the related independent work in \cite{fiderer2020general}.

\begin{acknowledgments}
	\textbf{Funding}. This work was supported by the National Key Research and Development Program of China (Grant Nos. 2017YFA0303703 and 2018YFA030602) and the National Natural Science Foundation of China (Grant Nos. 91836303, 61975077, 61490711 and 11690032) and Fundamental Research Funds for the CentralUniversities (Grant No. 020214380068). 
\end{acknowledgments}

\textbf{Disclosures}. The authors declare no conflicts of interest.
\bibliographystyle{unsrt}

\begin{thebibliography}{}

\end{thebibliography}


\begin{thebibliography}{10}

		
		\bibitem{r4}
		Max Born and Emil Wolf.
		\newblock Principles of optics, 7th (expanded) edition.
		\newblock {\em United Kingdom: Press Syndicate of the University of Cambridge},
		461, 1999.
		
		\bibitem{rayleigh1879xxxi}
		Lord Rayleigh.
		\newblock Xxxi. investigations in optics, with special reference to the
		spectroscope.
		\newblock {\em The London, Edinburgh, and Dublin Philosophical Magazine and
			Journal of Science}, 8(49):261--274, 1879.
		
		\bibitem{Hell:94}
		Stefan~W. Hell and Jan Wichmann.
		\newblock Breaking the diffraction resolution limit by stimulated emission:
		stimulated-emission-depletion fluorescence microscopy.
		\newblock {\em Opt. Lett.}, 19(11):780--782, Jun 1994.
		
		\bibitem{Heintzmann:02}
		Rainer Heintzmann, Thomas~M. Jovin, and Christoph Cremer.
		\newblock Saturated patterned excitation microscopy---a concept for optical
		resolution improvement.
		\newblock {\em J. Opt. Soc. Am. A}, 19(8):1599--1609, Aug 2002.
		
		\bibitem{Gustafsson13081}
		Mats G.~L. Gustafsson.
		\newblock Nonlinear structured-illumination microscopy: Wide-field fluorescence
		imaging with theoretically unlimited resolution.
		\newblock {\em Proceedings of the National Academy of Sciences},
		102(37):13081--13086, 2005.
		
		\bibitem{RN357}
		Michael~J. Rust, Mark Bates, and Xiaowei Zhuang.
		\newblock Sub-diffraction-limit imaging by stochastic optical reconstruction
		microscopy (storm).
		\newblock {\em Nature Methods}, 3(10):793--796, 2006.
		
		\bibitem{Betzig1642}
		Eric Betzig, George~H. Patterson, Rachid Sougrat, O.~Wolf Lindwasser, Scott
		Olenych, Juan~S. Bonifacino, Michael~W. Davidson, Jennifer
		Lippincott-Schwartz, and Harald~F. Hess.
		\newblock Imaging intracellular fluorescent proteins at nanometer resolution.
		\newblock {\em Science}, 313(5793):1642--1645, 2006.
		
		\bibitem{PhysRevLett.123.143604}
		L.~A. Howard, G.~G. Gillett, M.~E. Pearce, R.~A. Abrahao, T.~J. Weinhold,
		P.~Kok, and A.~G. White.
		\newblock Optimal imaging of remote bodies using quantum detectors.
		\newblock {\em Phys. Rev. Lett.}, 123:143604, Sep 2019.
		
		\bibitem{multiparameter}
		Magdalena Szczykulska, Tillmann Baumgratz, and Animesh Datta.
		\newblock Multi-parameter quantum metrology.
		\newblock {\em Advances in Physics: X}, 1(4):621--639, 2016.
		
		\bibitem{Matsumoto_2002}
		K~Matsumoto.
		\newblock A new approach to the cram{\'{e}}r-rao-type bound of the pure-state
		model.
		\newblock {\em Journal of Physics A: Mathematical and General},
		35(13):3111--3123, mar 2002.
		
		\bibitem{PhysRevLett.119.130504}
		Luca Pezz\`e, Mario~A. Ciampini, Nicol\`o Spagnolo, Peter~C. Humphreys, Animesh
		Datta, Ian~A. Walmsley, Marco Barbieri, Fabio Sciarrino, and Augusto Smerzi.
		\newblock Optimal measurements for simultaneous quantum estimation of multiple
		phases.
		\newblock {\em Phys. Rev. Lett.}, 119:130504, Sep 2017.
		
		\bibitem{PhysRevA.88.040102}
		O.~Pinel, P.~Jian, N.~Treps, C.~Fabre, and D.~Braun.
		\newblock Quantum parameter estimation using general single-mode gaussian
		states.
		\newblock {\em Phys. Rev. A}, 88:040102, Oct 2013.
		
		\bibitem{vidrighin2014joint}
		Mihai~D Vidrighin, Gaia Donati, Marco~G Genoni, Xian-Min Jin, W~Steven
		Kolthammer, MS~Kim, Animesh Datta, Marco Barbieri, and Ian~A Walmsley.
		\newblock Joint estimation of phase and phase diffusion for quantum metrology.
		\newblock {\em Nature communications}, 5:3532, 2014.
		
		\bibitem{PhysRevLett.111.070403}
		Peter~C. Humphreys, Marco Barbieri, Animesh Datta, and Ian~A. Walmsley.
		\newblock Quantum enhanced multiple phase estimation.
		\newblock {\em Phys. Rev. Lett.}, 111:070403, Aug 2013.
		
		\bibitem{r15}
		Philip J.~D. Crowley, Animesh Datta, Marco Barbieri, and I.~A. Walmsley.
		\newblock Tradeoff in simultaneous quantum-limited phase and loss estimation in
		interferometry.
		\newblock {\em Phys. Rev. A}, 89:023845, Feb 2014.
		
		\bibitem{r16}
		Sammy Ragy, Marcin Jarzyna, and Rafa\l{}
		Demkowicz-Dobrza\ifmmode~\acute{n}\else \'{n}\fi{}ski.
		\newblock Compatibility in multiparameter quantum metrology.
		\newblock {\em Phys. Rev. A}, 94:052108, Nov 2016.
		
		\bibitem{r13}
		Samuel~L. Braunstein and Carlton~M. Caves.
		\newblock Statistical distance and the geometry of quantum states.
		\newblock {\em Phys. Rev. Lett.}, 72:3439--3443, May 1994.
		
		\bibitem{giovannetti2011advances}
		Vittorio Giovannetti, Seth Lloyd, and Lorenzo Maccone.
		\newblock Advances in quantum metrology.
		\newblock {\em Nature photonics}, 5(4):222, 2011.
		
		\bibitem{PhysRevLett.96.010401}
		Vittorio Giovannetti, Seth Lloyd, and Lorenzo Maccone.
		\newblock Quantum metrology.
		\newblock {\em Phys. Rev. Lett.}, 96:010401, Jan 2006.
		
		\bibitem{Liu_2019}
		Jing Liu, Haidong Yuan, Xiao-Ming Lu, and Xiaoguang Wang.
		\newblock Quantum fisher information matrix and multiparameter estimation.
		\newblock {\em Journal of Physics A: Mathematical and Theoretical},
		53(2):023001, dec 2019.
		
		\bibitem{PhysRevA.96.012310}
		Haidong Yuan and Chi-Hang~Fred Fung.
		\newblock Quantum metrology matrix.
		\newblock {\em Phys. Rev. A}, 96:012310, Jul 2017.
		
		\bibitem{Chen_2017}
		Yu~Chen and Haidong Yuan.
		\newblock Maximal quantum fisher information matrix.
		\newblock {\em New Journal of Physics}, 19(6):063023, jun 2017.
		
		\bibitem{r5}
		Mankei Tsang, Ranjith Nair, and Xiao-Ming Lu.
		\newblock Quantum theory of superresolution for two incoherent optical point
		sources.
		\newblock {\em Phys. Rev. X}, 6:031033, Aug 2016.
		
		\bibitem{r6}
		Martin Pa\'{u}r, Bohumil Stoklasa, Zdenek Hradil, Luis~L. S\'{a}nchez-Soto, and
		Jaroslav Rehacek.
		\newblock Achieving the ultimate optical resolution.
		\newblock {\em Optica}, 3(10):1144--1147, Oct 2016.
		
		\bibitem{PhysRevA.97.023830}
		Mankei Tsang.
		\newblock Subdiffraction incoherent optical imaging via spatial-mode
		demultiplexing: Semiclassical treatment.
		\newblock {\em Phys. Rev. A}, 97:023830, Feb 2018.
		
		\bibitem{Tsang_2017}
		Mankei Tsang.
		\newblock Subdiffraction incoherent optical imaging via spatial-mode
		demultiplexing.
		\newblock {\em New Journal of Physics}, 19(2):023054, feb 2017.
		
		\bibitem{PhysRevLett.118.070801}
		Weng-Kian Tham, Hugo Ferretti, and Aephraim~M. Steinberg.
		\newblock Beating rayleigh's curse by imaging using phase information.
		\newblock {\em Phys. Rev. Lett.}, 118:070801, Feb 2017.
		
		\bibitem{Bonsma_Fisher_2019}
		Kent A~G Bonsma-Fisher, Weng-Kian Tham, Hugo Ferretti, and Aephraim~M
		Steinberg.
		\newblock Realistic sub-rayleigh imaging with phase-sensitive measurements.
		\newblock {\em New Journal of Physics}, 21(9):093010, sep 2019.
		
		\bibitem{grace2019approaching}
		Michael~R Grace, Zachary Dutton, Amit Ashok, and Saikat Guha.
		\newblock Approaching quantum limited super-resolution imaging without prior
		knowledge of the object location.
		\newblock {\em arXiv preprint arXiv:1908.01996}, 2019.
		
		\bibitem{PhysRevLett.121.250503}
		Micha\l{} Parniak, Sebastian Bor\'owka, Kajetan Boroszko, Wojciech Wasilewski,
		Konrad Banaszek, and Rafa\l{} Demkowicz-Dobrza\ifmmode~\acute{n}\else
		\'{n}\fi{}ski.
		\newblock Beating the rayleigh limit using two-photon interference.
		\newblock {\em Phys. Rev. Lett.}, 121:250503, Dec 2018.
		
		\bibitem{mqmt}
		J.~\ifmmode \check{R}\else \v{R}\fi{}eha\ifmmode~\check{c}\else \v{c}\fi{}ek,
		Z.~Hradil, B.~Stoklasa, M.~Pa\'ur, J.~Grover, A.~Krzic, and L.~L.
		S\'anchez-Soto.
		\newblock Multiparameter quantum metrology of incoherent point sources: Towards
		realistic superresolution.
		\newblock {\em Phys. Rev. A}, 96:062107, Dec 2017.
		
		\bibitem{PhysRevA.98.012103}
		J.~\ifmmode \check{R}\else \v{R}\fi{}eh\'a\ifmmode~\check{c}\else \v{c}\fi{}ek,
		Z.~Hradil, D.~Koutn\'y, J.~Grover, A.~Krzic, and L.~L. S\'anchez-Soto.
		\newblock Optimal measurements for quantum spatial superresolution.
		\newblock {\em Phys. Rev. A}, 98:012103, Jul 2018.
		
		\bibitem{Prasad_2020}
		Sudhakar Prasad.
		\newblock Quantum limited super-resolution of an unequal-brightness source pair
		in three dimensions.
		\newblock {\em Physica Scripta}, 95(5):054004, mar 2020.
		
		\bibitem{r11}
		Evangelia Bisketzi, Dominic Branford, and Animesh Datta.
		\newblock Quantum limits of localisation microscopy.
		\newblock {\em New Journal of Physics}, 21(12):123032, dec 2019.
		
		\bibitem{r7}
		Yiyu Zhou, Jing Yang, Jeremy~D. Hassett, Seyed Mohammad~Hashemi Rafsanjani,
		Mohammad Mirhosseini, A.~Nick Vamivakas, Andrew~N. Jordan, Zhimin Shi, and
		Robert~W. Boyd.
		\newblock Quantum-limited estimation of the axial separation of two incoherent
		point sources.
		\newblock {\em Optica}, 6(5):534--541, May 2019.
		
		\bibitem{PhysRevLett.121.180504}
		Zhixian Yu and Sudhakar Prasad.
		\newblock Quantum limited superresolution of an incoherent source pair in three
		dimensions.
		\newblock {\em Phys. Rev. Lett.}, 121:180504, Oct 2018.
		
		\bibitem{PhysRevA.99.022116}
		Sudhakar Prasad and Zhixian Yu.
		\newblock Quantum-limited superlocalization and superresolution of a source
		pair in three dimensions.
		\newblock {\em Phys. Rev. A}, 99:022116, Feb 2019.
		
		\bibitem{r9}
		Carmine Napoli, Samanta Piano, Richard Leach, Gerardo Adesso, and Tommaso
		Tufarelli.
		\newblock Towards superresolution surface metrology: Quantum estimation of
		angular and axial separations.
		\newblock {\em Phys. Rev. Lett.}, 122:140505, Apr 2019.
		
		\bibitem{r18}
		Mikael~P. Backlund, Yoav Shechtman, and Ronald~L. Walsworth.
		\newblock Fundamental precision bounds for three-dimensional optical
		localization microscopy with poisson statistics.
		\newblock {\em Phys. Rev. Lett.}, 121:023904, Jul 2018.
		
		\bibitem{Larson:18}
		Walker Larson and Bahaa E.~A. Saleh.
		\newblock Resurgence of rayleigh's curse in the presence of partial coherence.
		\newblock {\em Optica}, 5(11):1382--1389, Nov 2018.
		
		\bibitem{Tsang:19}
		Mankei Tsang and Ranjith Nair.
		\newblock Resurgence of rayleigh's curse in the presence of partial coherence:
		comment.
		\newblock {\em Optica}, 6(4):400--401, Apr 2019.
		
		\bibitem{Larson:19}
		Walker Larson and Bahaa E.~A. Saleh.
		\newblock Resurgence of rayleigh's curse in the presence of partial coherence:
		reply.
		\newblock {\em Optica}, 6(4):402--403, Apr 2019.
		
		\bibitem{Hradil:19}
		Zden\v{e}k Hradil, Jaroslav \v{R}eh\'{a}\v{c}ek, Luis S\'{a}nchez-Soto, and
		Berthold-Georg Englert.
		\newblock Quantum fisher information with coherence.
		\newblock {\em Optica}, 6(11):1437--1440, Nov 2019.
		
		\bibitem{r19}
		Yoav Shechtman, Steffen~J. Sahl, Adam~S. Backer, and W.~E. Moerner.
		\newblock Optimal point spread function design for 3d imaging.
		\newblock {\em Phys. Rev. Lett.}, 113:133902, Sep 2014.
		
		\bibitem{nehme2020learning}
		Elias Nehme, Boris Ferdman, Lucien~E Weiss, Tal Naor, Daniel Freedman, Tomer
		Michaeli, and Yoav Shechtman.
		\newblock Learning an optimal psf-pair for ultra-dense 3d localization
		microscopy.
		\newblock {\em arXiv preprint arXiv:2009.14303}, 2020.
		
		\bibitem{PhysRevLett.123.193601}
		J.~\ifmmode \check{R}\else \v{R}\fi{}eh\'a\ifmmode~\check{c}\else \v{c}\fi{}ek,
		M.~Pa\'ur, B.~Stoklasa, D.~Koutn\'y, Z.~Hradil, and L.~L. S\'anchez-Soto.
		\newblock Intensity-based axial localization at the quantum limit.
		\newblock {\em Phys. Rev. Lett.}, 123:193601, Nov 2019.
		
		\bibitem{r17}
		Martin Pa\'{u}r, Bohumil Stoklasa, Jai Grover, Andrej Krzic, Luis~L.
		S\'{a}nchez-Soto, Zden\v{e}k Hradil, and Jaroslav \v{R}eh\'{a}\v{c}ek.
		\newblock Tempering rayleigh's curse with psf shaping.
		\newblock {\em Optica}, 5(10):1177--1180, Oct 2018.
		
		\bibitem{Koutn2021}
		D~Koutn{\'{y}}, Z~Hradil, J~{\v{R}}eh{\'{a}}{\v{c}}ek, and L~L
		S{\'{a}}nchez-Soto.
		\newblock Axial superlocalization with vortex beams.
		\newblock {\em Quantum Science and Technology}, 6(2):025021, mar 2021.
		
		\bibitem{helstrom1976quantum}
		Carl~W Helstrom and Carl~W Helstrom.
		\newblock {\em Quantum detection and estimation theory}, volume~3.
		\newblock Academic press New York, 1976.
		
		\bibitem{holevo2011probabilistic}
		Alexander~S Holevo.
		\newblock {\em Probabilistic and statistical aspects of quantum theory},
		volume~1.
		\newblock Springer Science \& Business Media, 2011.
		
		\bibitem{Carollo_2019}
		Angelo Carollo, Bernardo Spagnolo, Alexander~A Dubkov, and Davide Valenti.
		\newblock On quantumness in multi-parameter quantum estimation.
		\newblock {\em Journal of Statistical Mechanics: Theory and Experiment},
		2019(9):094010, sep 2019.
		
		\bibitem{PhysRevX.10.031023}
		Mankei Tsang, Francesco Albarelli, and Animesh Datta.
		\newblock Quantum semiparametric estimation.
		\newblock {\em Phys. Rev. X}, 10:031023, Jul 2020.
		
		\bibitem{datta2020sub}
		Chandan Datta, Marcin Jarzyna, Yink~Loong Len, Karol {\L}ukanowski, Jan
		Ko{\l}ody{\'n}ski, and Konrad Banaszek.
		\newblock Sub-rayleigh resolution of incoherent sources by array homodyning.
		\newblock {\em arXiv preprint arXiv:2005.08693}, 2020.
		
		\bibitem{doi:10.1142/S0219749919410156}
		Yink~Loong Len, Chandan Datta, MichaÅ,Parniak, and Konrad Banaszek.
		\newblock Resolution limits of spatial mode demultiplexing with noisy
		detection.
		\newblock {\em International Journal of Quantum Information}, 18(01):1941015,
		2020.
		
		\bibitem{fiderer2020general}
		Lukas~J Fiderer, Tommaso Tufarelli, Samanta Piano, and Gerardo Adesso.
		\newblock General expressions for the quantum fisher information matrix with
		applications to discrete quantum imaging.
		\newblock {\em arXiv preprint arXiv:2012.01572}, 2020.

	


	
	
\end{thebibliography}

\end{document}